\documentclass{jnmp} 
 
\usepackage{amsmath} 
\usepackage{epsfig} 
 
\JNMPnumberwithin{equation}{section} 
 
\theoremstyle{definition}

\begin{document} 
 
%
\renewcommand{\evenhead}{A Krylovas and R \v{C}iegis} 
\renewcommand{\oddhead}{Asymptotic Approximation of Hyperbolic Weakly Nonlinear Systems} 
 
%
\thispagestyle{empty} 
 
\FirstPageHead{8}{4}{2001}{\pageref{krylovas-firstpage}--\pageref{krylovas-lastpage}}{Letter} 
 
\copyrightnote{2001}{A Krylovas and R \v{C}iegis} 
 
\Name{Asymptotic Approximation\\ of Hyperbolic Weakly Nonlinear Systems} 
\label{krylovas-firstpage} 
 
\Author{A KRYLOVAS and R \v{C}IEGIS} 
 
\Address{Vilnius Gediminas Technical University, Saul{\. e}tekio al. 11, 2054, Vilnius, Lithuania \\  
E-mail: akr@fm.vtu.lt, \ rc@fm.vtu.lt} 
 
\Date{Received April 11, 2001; Revised July 7, 2001; Accepted July 20, 
2001}

\begin{abstract} 
\noindent 
An averaging  method for getting uniformly valid  
asymptotic approximations of the solution of hyperbolic 
systems of equations is presented.  
The averaged system of equations  
disintegrates into  
independent equations for non-resonance systems. We consider  
the resonance conditions for some classes of solutions.  
The averaged system can be solved numerically in the resonance case.  
The shallow water prob\-lem is considered 
as an example of the resonance system. Results of numerical experiments 
are presented. 
\end{abstract}

\section{Introduction} 
 
In this paper we consider a system of weakly nonlinear equations with a small 
positive parameter $\varepsilon$: 
\begin{equation} 
\label{math/1} 
U_t+A(U)U_x=\varepsilon B(t,x,{\varepsilon}t,{\varepsilon}x,U,U_x,U_{xx},U_{xxx}), 
\end{equation} 
where $U=(u_1,u_2,\ldots,u_n)^T$ is a column vector,  
$A =\left\|a_{ij}\right\|$ is an $n\times n$ matrix, 
and $B=(b_1,b_2,\ldots,b_n)^T$ is a column vector.  
We assume that all coefficients are sufficiently smooth functions. 
 
Many physical problems are described by such systems. 
We mention only some examples: dispersive waves in plasma,  
problems of weakly nonlinear optics, one-dimensional gas dynamic equations, shallow 
water waves.  
The construction of uniform asymptotic approximations of the 
solution of (\ref{math/1}) becomes a nontrivial task if secular terms 
arise in the expansion. We can illustrate this situation by 
considering the following linear problem 
\begin{gather} 
   u_t + u_x = \varepsilon u , \nonumber \\ 
   u(0, x) = \sin x .\label{i1}  
\end{gather} 
We find easily the exact solution of problem (\ref{i1}) 
\[ 
   u(t, x, \varepsilon) = e^{\varepsilon t} \sin(x-t) . 
\] 
Using a Taylor series expansion of 
the function $e^{\varepsilon t}$  
we get that 
\begin{equation} 
   u(t, x, \varepsilon) = sin(x-t) + \varepsilon t \sin(x-t) + 
         \frac{\varepsilon^2 t^2}{2} \sin(x-t) + \cdots . 
                                 \label{i2} 
\end{equation} 
This expansion has secular terms $\varepsilon t$, $\varepsilon^2 t^2$, 
$\ldots$, and therefore the formula (\ref{i2}) is asymptotical only 
if $\varepsilon t \ll 1$. 
In this case we have that  
\[ 
    u(t, x, \varepsilon) = \sin(x-t) + O(\varepsilon t). 
\] 
 
For $\varepsilon t = O(1)$ this expansion is not asymptotical.  
On the other hand the problem (\ref{i1}) has a classical solution in 
any domain 
\[ 
    0 < t + |x| < c_0 / \varepsilon , 
\] 
here $c_0$ is a constant. It is easy to prove that 
\[ 
   \lim_{\varepsilon \rightarrow 0} u(t, x, \varepsilon)  \not= \sin(x-t) . 
\] 
Therefore it is a nontrivial task to 
construct an asymptotical approximation, which is 
uniformly valid in the region $t + |x| = O\left(\varepsilon^{-1}\right)$.  
 
The basic idea of all asymptotic methods is to introduce 
new ``slow'' variables, e.g., $\tau=\varepsilon t$, 
$\xi=\varepsilon x$, and to define explicitly the dependence on 
``fast'' variables. For example in \cite{Tan} the solution is obtained 
in the following form:  
\begin{gather*} 
u_j=\psi_j(\zeta,\eta_j) + O(\varepsilon) , \qquad      
\zeta=\varepsilon^{1+a}t, \qquad 
\eta_j=\varepsilon^a\left(x-\lambda_jt+\varepsilon^{1-a}\varphi(t,x)\right).   
\end{gather*} 
Substituting these expressions into system (\ref{math/1}),  
using a  
Taylor expansion with respect to $\varepsilon$, and equating    
coefficients at $\varepsilon^j$ 
we get equations for new unknown functions $\psi_j$ 
\[ 
  \psi_{j\zeta}+(\alpha_j\psi_j+\beta_j)\psi_{j\eta_j}+ 
  \gamma_j\psi_{j\eta_j\eta_j\eta_j}+\delta_j\psi_j = 0 . 
\] 
 
Our goal is to approximate the initial nonlinear problem 
by some fundamental equations, for which the analytical solution exists.  
The Burgers' equation and the Korteweg-de Vries equations are examples of such 
problems. Most asympto\-ti\-cal methods 
are based on  
physical assumptions and no strict mathematical proofs  
of the validity of these approximations are given 
\cite{Bhat, Jef}. 
Important applications of asymptotic methods are given in 
\cite{chikwendu, florence, skinner}.  
A survey of mathematical results on asymptotic expansion methods   
is presented by Kalyakin in \cite{Kal1}, see also Keworkian and Cole  
\cite{Cole}. 
Similar problems arise not only for systems of 
equations, but also for semilinear perturbed wave 
equations, the telegraph equation and weakly nonlinear beam  
equation \cite{boer, easwaran, lai}.  
 
Usually application of  
formal asymptotic method reduces the initial problem  
to a single nonlinear wave 
equation \cite{Tan}. Our method reduces problem (\ref{math/1}) to the  
integro-differential system of averaged 
equations. A new problem can be seen as a more difficult problem,  
than the initial formulation. 
However, we will show that the averaged system disintegrates into independent 
equations in the non-resonance case. Moreover, our method allows  
to describe accurately the 
resonance interaction of waves, as well. 
Our internal averaging method is close to the multiple-scale method, 
which is used for similar problems in \cite{easwaran}.   
The application of internal averaging method for solving gas dynamic  
equations is presented in \cite{ciegis}. 
 
The rest of the paper is organized as follows. 
In Section~\ref{sec/1}, we formulate the general averaging scheme. 
We propose the averaging operators and  
compare our scheme with the other similar methods.  
In Section~\ref{sec/2}, we consider the resonance case, 
when the averaged system is connected.  
Section~\ref{sec/3} deals with the application of the proposed  
algorithm for a shallow water model.  
It is proved that the initial system is ill-posed, hence the 
regularized model is formulated. The averaged system also gives  
a nontrivial regularization of the shallow water model. Finally we give 
the results of numerical experiments. 
 
\section{Method of averaging} \label{sec/1} 
 
Let $U_0$ be a constant solution of equation (\ref{math/1}):  
\begin{equation} 
\label{math/2} 
B(t,x,{\varepsilon}t,{\varepsilon}x,U_0,0,0,0)=0. 
\end{equation} 
We assume that problem (\ref{math/1}) is hyperbolic  
in the neighborhood of $U_0$, i.e. 
there exists an    
$n\times n$ matrix $R=\left\|r_{ij}\right\|$ $\det R \ne 0$ such that 
\begin{equation} 
\label{math/3} 
\Lambda \equiv {\rm diag}\, \{\lambda_1,\lambda_2,\ldots,\lambda_n\}=R A(U_0) R^{-1}. 
\end{equation} 
We are interested in finding a small-amplitude wave solution 
\begin{equation} 
\label{math/4} 
U(t,x,\varepsilon)=U_0+\varepsilon U_1(t,x,\varepsilon). 
\end{equation} 
Linearizing equations (\ref{math/1})  
with respect to $U_0$ yields the system of equations   
\begin{equation} 
\label{math/5} 
V_t+\Lambda V_z= 
\varepsilon F(t,x,\varepsilon t,\varepsilon x,V,V_x,V_{xx},V_{xxx})+o(\varepsilon), 
\end{equation} 
where  
$V=RU_1$, and the function $F$ is given by    
\begin{gather*} 
F = -RA_1\left[R^{-1}V\right]R^{-1}V_x+R \left(B_0\left[R^{-1}V\right]+B_1\left[R^{-1}V_x\right] \right.\\ 
\left.\qquad{} + B_2\left[R^{-1}V_{xx}\right]+B_3\left[R^{-1}V_{xxx}\right]\right), \\ 
A_1[U_1]\equiv\frac{dA(U_0)}{dU}U_1=\left\|\sum\limits_{k=1}^n\left[ 
\frac{\partial}{\partial u_k} a_{ij}(U_0)\right]u_{1k}\right\|,\\ 
B_m[U_1]=(b_{m1},b_{m2},\ldots,b_{mn})^T, \qquad  m=0,1,2,3,\\ 
b_{0j} =\sum\limits_{k=1}^n\frac{\partial b_j(t,x,\varepsilon t,  
\varepsilon x,U_0,0,0,0)}{\partial u_{0k} }u_{1k} , \qquad 
b_{1j} = \sum\limits_{k=1}^n\frac{\partial b_j}{\partial u_{0kx} }u_{1kx}, \\ 
b_{2j} = \sum\limits_{k=1}^n\frac{\partial b_j}{\partial u_{0kxx} }u_{1kxx},  
\qquad 
b_{3j} = \sum\limits_{k=1}^n\frac{\partial b_j}{\partial u_{0kxxx} }u_{1kxxx}. 
\end{gather*} 
 
If $\varepsilon=0$, then system (\ref{math/5}) with the initial condition 
\begin{equation} 
\label{math/6} 
V(0,x,\varepsilon)=V_0(x) 
\end{equation} 
describes $n$ independent linear waves $v_j = v_{0j}(x-\lambda_jt)$. 
 
If $t+|x|\sim \varepsilon^{-1}$, then the exact solution of  
initial-value problem 
(\ref{math/5}), (\ref{math/6}) 
(and therefore also  of problem (\ref{math/1})) is not close to   
the simple wave.  
For example, the nonlinear equation  
\[ 
v_t+v_x=\varepsilon v v_x 
\] 
describes a nonlinear wave, which is given by the implicit relation  
\[ 
 v(t,x,\varepsilon)=v_0(x-t+\varepsilon t v(t,x)) . 
\] 
Obviously it can not be approximated by a simple wave $v=v_0(x-t)$  
for  $\varepsilon t=O(1)$. 
 
We first rewrite system (\ref{math/5}), (\ref{math/6}) in a coordinate form 
\begin{gather} 
 \frac{\partial v_j}{\partial t}+\lambda_j\frac{\partial v_j}{\partial x}  
 = \varepsilon f_j \left(t,x,\varepsilon t, \varepsilon x, V,  
 \frac{\partial V}{\partial x}, 
 \frac{\partial^2 V}{\partial x^2},\frac{\partial^3 V}{\partial x^3}\right),  \nonumber\\ 
v_j(0,x,\varepsilon) = v_{0j}(\varepsilon x, x), \qquad j=1,2,\ldots,n.\label{math/7} 
\end{gather} 
 
Let $\tau=\varepsilon t$, $\xi=\varepsilon x$ be ``slow'' variables, 
and $y_j=x-\lambda_jt$, $j=1,2,\ldots,n$ be  
``fast'' characteristic variables. The operator of averaging  
along the $j$-th characteristic of the 
non-perturbed (i.e., $\varepsilon=0$) system (\ref{math/7}) is given by  
\begin{gather} 
M_j[g(t,x,\tau,\xi,v_1,v_2,\ldots,v_n)] \equiv  
\lim_{T\rightarrow \infty}\frac{1}{T}\int_0^Tg\Big(s,y_j+\lambda_js, \tau,  
  \xi,  \nonumber \\ 
   \qquad v_1(y_j+(\lambda_j-\lambda_1)s), \ldots, 
  v_n(y_j+(\lambda_j-\lambda_n)s)\Big)ds . \label{math/8}  
\end{gather} 
 
We define the following averaged system of equations 
\begin{gather} 
\frac{\partial w_j}{\partial\tau}+\lambda_j\frac{\partial w_j}{\partial \xi}= 
M_j\left[f_j\left(t,x,\tau,\xi,W,\frac{\partial W}{\partial y_k}, 
\frac{\partial^2 W}{\partial y_k^2},\frac{\partial^3 W} 
             {\partial y_k^3}\right)\right] , \nonumber   \\ 
  w_j(0,\xi,y_j) = v_{0j}(\xi,y_j), \qquad  j=1,2,\ldots,n.\label{math/9} 
\end{gather} 
 
If the operator 
$B \equiv 0$, then averaged system (\ref{math/9}) takes the form 
\begin{equation} 
\label{math/10} 
\frac{\partial w_j}{\partial\tau}+\lambda_j\frac{\partial w_j}{\partial \xi}= 
\sum\limits_{k=1}^n\sum\limits_{m=1}^nf_{jkm}M_j 
    \left[w_k\frac{\partial w_m}{\partial y_m}\right]. 
\end{equation} 
 
Without a loss of generality we can assume, that 
\begin{equation} 
\label{math/11} 
\lim_{T \rightarrow \infty}\frac{1}{2T}\int_ 
{-T}^{T}v_{0j}(\xi,x)dx=0. 
\end{equation} 
It is easy to prove that the following properties of the averaging  
operators $M_j$ are valid: 
\begin{gather} 
M_j\left[v_j\frac{\partial v_j}{\partial y_j}\right] \equiv 
 v_j \frac{\partial v_j}{\partial y_j}, \qquad 
  M_j\left[v_i\frac{\partial v_j}{\partial y_j}\right] \equiv  
 M_j[v_i] \frac{\partial v_j}{\partial y_j},\nonumber \\ 
M_j\left[v_j\frac{\partial v_i}{\partial y_i}\right] \equiv 0, \qquad  i \ne j.\label{math/12} 
 \end{gather} 
 
For $n=2$ 
using (\ref{math/10})--(\ref{math/12}) we get 
that the averaged system of 
equations reduces to two independent problems: 
\begin{gather} 
\frac{\partial w_j}{\partial\tau}+\lambda_j\frac{\partial w_j}{\partial \xi}= 
f_{jjj}w_j\frac{\partial w_j}{\partial y_j},  \nonumber \\  
   w_j(0,\xi,y_j) = v_{0j}(\xi,y_j).\label{math/13} 
 \end{gather} 
 
If the operator $B$ 
is a linear function of $U_{xx}$ or $U_{xxx}$, then averaged equations 
(\ref{math/13}) are described by the 
Burgers' or the Korteweg-de Vries equations. 
 
In most cases equations of averaged system 
(\ref{math/9}) are connected and they describe the 
interaction of waves.   
Asymptotic analysis is finished at this stage, but some 
numerical analysis is still needed in order to get the solution. 
 
\section{Approximation accuracy analysis} \label{sec/2} 
Let $\mathcal{M}$ be a class of functions, for which averages (\ref{math/8}) 
exist uniformly for all variables.  
Then the solution of averaged system (\ref{math/7}) exists 
and we can analyze the approximation properties of this solution.  
 
The case of 
periodical initial conditions $u_{0j}(x)\in C_{2\pi}^{1}(\mathbb{R})$ 
was considered in \cite{Kryl1,Star}.  
For sufficiently smooth functions 
$f_j(u_1,\ldots,u_n,u_{1x},\ldots,u_{nx})$ it was proved  
that if $(v_1,v_2,\ldots$, $v_n)$ 
is the solution of system (\ref{math/7}) and  
$(w_1,w_2,\ldots,w_n)$ is the solution 
of averaged system (\ref{math/9}), then there exists a constant 
$c_{0}>0$ such that  
\begin{equation} 
\label{math/14} 
\lim_{\varepsilon \rightarrow 0}\max_j\sup_{0\le t+|x|\le \frac{c_{0}} 
{\varepsilon}} 
|u_j(t,x,\varepsilon)-v_j(\varepsilon t, \varepsilon x, x - \lambda_j t)|=0. 
\end{equation} 
This result is analogous to the First Bogoliubov's theorem  
for ordinary differential equations \cite{Bog}. 
 
The essence of the proposed averaging method is the following: the ave\-rage 
$M_j[g(\tau,$ $\xi, t, x, v)]$ 
along the  $j$-th characteristic $y_j=x-\lambda_jt$ is the limit of the integral 
\[ 
 \lim_{T\rightarrow \infty}\frac{1}{T}\int_0^T 
g(\tau,\xi,s,y_j+\lambda_js,v(\tau,\xi,s,y_j+\lambda_js))ds. 
\] 
We integrate function $v$, which is still unknown,  
that is why our method can be 
called the \textit{internal} averaging. For comparison the classical averaging  
along characteristics \cite{Mitr} is defined by: 
\[ 
 \lim_{T\rightarrow \infty}\frac{1}{T}\int_0^T 
g(\tau,\xi,s,y_j+\lambda_js,v(\tau,\xi,t,x))ds 
\] 
and it can be called the \textit{external} averaging.  
 
We illustrate the difference between these two techniques, by  
considering   
the model system with the internal resonance 
\begin{gather*} 
 u_t+u_x = \varepsilon v \sin x,  \qquad u(0,x,\varepsilon)=0, \\ 
 v_t = 0, \qquad  v(0,x)=\sin x. 
\end{gather*} 
 
The \textit{external} averaging 
gives the problem  
\[ 
  U_t+U_x=0,\qquad U(0,x)=0 
\] 
and $U\equiv 0$ does not approximate the exact solution 
\[ 
 u(t,x,\varepsilon)=\frac{\varepsilon}{4}(2t+\sin 2(x-t)-\sin2x) 
\] 
if $t \sim\varepsilon^{-1}$. The \textit{internal} averaging of this system 
gives the problem 
\begin{gather*} 
 U_{\tau} = \lim\limits_{T \rightarrow \infty} \frac{1}{T} \int_0^T  
   V(\tau,y+s) \sin(y+s)ds, \qquad U(0,y)=0, \\ 
  V_{\tau}=0, \qquad V(0,x)=\sin x, \qquad y=x-t . 
\end{gather*} 
After simple computations we get 
\begin{gather*} 
  V(\tau,x) = \sin x , \\ 
  U_{\tau} = \lim_{T\rightarrow \infty}\frac{1}{T}\int_0^T 
      \sin^2(y+s)ds = \frac{1}{2}. 
\end{gather*} 
Hence $U=\frac{\varepsilon t}{2}$ and the equality $u=U+o(1)$ 
is satisfied uniformly for 
$t\in\left[0,O\left(\varepsilon^{-1}\right)\right]$. 
 
Using Fourier series we can write the resonance conditions for  
system (\ref{math/7}). 
Let $f_j(t,x,\tau,\xi,v_1,\ldots v_n)$ be periodic functions with respect 
to $t$: 
\begin{gather*} 
  f_j\left(t+\Lambda_j^t,x+\Lambda_j^x,\tau,\xi,v_1(\tau,\xi,y_1+\Lambda_1), 
  \ldots,v_ n(\tau,\xi,y_n+\Lambda_n)\right)  \\ 
\qquad {}  = f_j(t,x,\tau,\xi,v_1(\tau,\xi,y_1),\ldots,v_n( 
  \tau,\xi,y_n)).  
\end{gather*} 
Integrating these functions along characteristics we get that 
\begin{gather*} 
f_j(t,x,\tau,\xi,v_1(\tau,\xi,y_1),\ldots,v_n(\tau,\xi,y_n)) \\ 
 \qquad = \sum_{(l^t,l^x,l_1,\ldots,l_n)\ne0}f_{jl}(\tau,\xi)\, e^{2\pi i 
 \left(\frac{l^tt} {\Lambda^t_j}+\frac{l^xx}{\Lambda_j^x}+\frac{l_1y_1}  
  {\Lambda_1}+\cdots+\frac{l_ny_n}{\Lambda_n}\right)} . 
\end{gather*} 
Then using the substitution  
\[ 
   t=x, \quad x=y_j+\lambda_js, \quad y_i=y_j+(\lambda_j-\lambda_i)s 
\] 
we obtain that the condition 
\begin{equation} 
 \lim_{T\rightarrow \infty}\frac{1}{T}\int_0^Tf_j(s,y_j+\lambda_js, 
 \tau,\xi,\ldots y_j+(\lambda_j-\lambda_i)s\ldots)ds=0 
\label{math/25} 
\end{equation} 
is satisfied if and only if 
\begin{gather} 
\frac{l^t}{\Lambda_j^t}+ \frac{\lambda_j l^x}{\Lambda_j^x}+ 
\sum\limits_{k\ne j}\frac{\lambda_j-\lambda_k}{\Lambda_k}l_k \ne 0,  
\nonumber \\ 
 \forall \; l^t,l^x,l_k\in\mathbb{Z} \quad  \& \quad 
 |l^t|+|l^x|+\sum\limits_{k\ne j}|l_k|\ne 0.\label{math/15} 
\end{gather} 
The condition (\ref{math/25}) 
specifies the absence of the resonance in system (\ref{math/7}). 
If it is satisfied, then  averaged system (\ref{math/9}) 
disintegrates into independent equations, which are similar to (\ref{math/13}). 
 
It was proved in \cite{Kryl2}, 
that the class of functions $\mathcal{M}$, for which 
the \textit{internal} ave\-ra\-ging can be applied, includes almost 
periodic functions. 
For example, let assume that   
\begin{gather*} 
f_j=\rho_j(\tau,\xi, t, x)u_1u_2\cdots u_n, \qquad  
\rho_j=\sum_{l=(l^t,l^x)\ne0}\rho_j(\tau,\xi)e^{i(\nu^t_{jl^t}t+  
 \nu^x_{jl^x}x)}, \\  
u_{0j}\sim \sum_{k\in \mathbb{Z} } u_{jk}(\xi)e^{i\nu_{jk}^0x }. 
\end{gather*} 
Then system (\ref{math/7}) is non-resonance if 
\begin{gather} 
     \forall \; l^t,l^x,l_k\in\mathbb{Z} \qquad  
   |l^t|+|l^x|+\sum\limits_{k\ne j}|l_k|\ne 0: \nonumber\\  
    \nu_{jl^t}^t+\nu_{jl^x}^x\lambda_j+\sum\limits_{k\ne j} 
    \nu_{kl_k}^0(\lambda_j-\lambda_k)\ne 0 .  \label{math/16} 
\end{gather} 
 
Thus our method treats uniformly both the resonance and non-resonance 
problems. The mathematical justification of the method was given 
only in the case,  
when the operator~$B$ in system (\ref{math/1}) does not depend on $U_{xx}$ and 
$U_{xxx}$. 
But our results for non-resonance systems coincide with the results given 
by some other methods. Therefore we expect that asymptotic solution 
approximates uniformly the exact 
solution in more general cases too. 
The numerical analysis of such problems will be given in the next section.

\section{Shallow water waves} \label{sec/3} 
In this section we consider the system of    
shallow water equations \cite{Alesh} 
\begin{gather} 
Z_t+(HU)_x = \varepsilon \left(\frac{1}{6}\left(H^3U_{xx}\right)_x - \frac{1}{2} 
   (HU)_{xxx} - HH_x(HU)_{xx}-(ZU)_x \right) ,  \nonumber\\ 
U_t+Z_x = -\varepsilon U U_x, 
\label{math/18} 
\end{gather} 
where $z = \varepsilon Z$ is the water surface level, 
$u = \varepsilon U$ is the horizontal velocity of the fluid, 
$H$ is the normalized bottom equation. 
All variables are normalized by some typical horizontal ($L_*$) and 
vertical ($H_*$) sizes: 
\begin{gather*} 
  x = \frac{x_1}{L_*}, \qquad z=\frac{z_1}{H_*}, \qquad   
   t=\frac{\sqrt{gH_*}}{L_*}t_1, \qquad  
  H = \frac{H_1}{H_*}, \quad \varepsilon =\left(\frac{H_*}{L_*}\right)^2  
                    \ll 1 , 
\end{gather*} 
where $x_1$ is the horizontal coordinate, $t_1$ the time,  
$z_1$ the water surface equation, $g$ the acceleration due to gravity, $H_1(x)$ 
the bottom equation. 
We assume, that 
\[ 
 H=1+\varepsilon h(x). 
\] 
Then we can simplify the first equation of the system:  
\begin{gather} 
 Z_t+U_x = \varepsilon \left( -\frac{1}{3}U_{xxx}-(h U)_x-(Z U)_x \right), \nonumber\\  
  U_t+Z_x = -\varepsilon U U_x .\label{math/19} 
\end{gather} 
We define new functions $v^+$ and $v^-$, which are related to $U$ and $Z$  
in the following way: 
\[ 
 U=v^+-v^-,\qquad Z=v^++v^-. 
\] 
Then problem (\ref{math/19}) can be reduced to the system 
\begin{gather} 
v^+_t+v^+_x = -\dfrac{\varepsilon}{2}\Bigg(\frac{1}{3}\left(v^+_{xxx}-v^-_{xxx}\right)+ 
\left( h(x)\left(v^+-v^-\right) \right)_x  \nonumber\\  
\qquad {}+ \left(\left(v^+\right)^2-\left(v^-\right)^2\right)_x+\left(v^+-v^-\right) 
\left(v^+_x-v^-_x\right)\Bigg), \nonumber\\  
v^-_t-v^-_x = -\frac{\varepsilon}{2}\Bigg(\frac{1}{3}\left(v^+_{xxx}-v^-_{xxx}\right)+ 
    \left(h(x)\left(v^+-v^-\right)\right)_x  \nonumber \\ 
  \qquad {}- \left(\left(v^+\right)^2-\left(v^-\right)^2\right)_x 
-\left(v^+-v^-\right)\left(v^+_x-v^-_x\right)\Bigg) . \label{math/20}  
\end{gather} 
 
The asymptotic solution 
$V^{\pm}(\tau,y^{\pm}) = v^{\pm}(t,x,\varepsilon) + o(1)$,  $y^{\pm}=x\mp t$ 
satisfies the averaged system 
\begin{gather} 
\frac{\partial V^+}{\partial \tau} = -\frac{1}{2}\left(\frac{1}{3}V^+_{y^+y^+y^+}+ 
\left(\left\langle h(x)V^-\right\rangle_+\right)_{y^+}+3V^+V^+_{y^+}\right), 
\nonumber \\  
\frac{\partial V^-}{\partial \tau} = \frac{1}{2}\left(\frac{1}{3}V^-_{y^-y^-y^-}+ 
\left(\left\langle h(x)V^+\right\rangle_-\right)_{y^-}+3V^-V^-_{y^-}\right),\label{math/21}  
\end{gather} 
here we have denoted the averaging operators: 
\begin{gather*} 
\left\langle h(x)V^+\right\rangle_-=\lim_{T\rightarrow \infty}\frac{1}{T} 
\int_0^Th(y^--s)V^+(\tau,y^--2s)ds , \\ 
\left\langle h(x)V^-\right\rangle_+=\lim_{T\rightarrow \infty}\frac{1}{T} 
\int_0^Th(y^++s)V^-(\tau,y^++2s)ds. 
\end{gather*} 
 
In \cite{Alesh} the analysis of the same problem  
is based on the assumption that $U = Z$, i.e. only one wave 
is considered. After simple computations we  get the 
following Korteweg-de Vries problem for $Z$: 
\[ 
  Z_{t} + Z_x + \frac{3}{2} \varepsilon Z Z_x 
  + \frac{1}{6} \varepsilon Z_{xxx} = 0 . 
\] 
We will prove that in the non-resonance case 
system (\ref{math/21}) describes two independent waves. 
 
Let assume, that (\ref{math/11}) holds for the initial condition,  
then we get 
\[ 
\left\langle V^{\pm}\right\rangle_{\mp}=\lim_{t\rightarrow \infty} 
\frac{1}{T}\int\limits_o^TV^{\pm}(\tau,y)dy=0 . 
\] 
System (\ref{math/20}) is non-resonance if the following  
equalities  
\begin{equation} 
\label{math/22} 
\langle h(x)V^-\rangle_+ = 0, \qquad \langle h(x)V^+\rangle_-=0 
\end{equation} 
hold. 
Then averaged system (\ref{math/21}) reduces to two Korteweg-de Vries 
equations: 
\[ 
\frac{\partial V^{\pm} }{\partial \tau}\pm\frac{3}{2}V^{\pm}\frac{\partial V^{\pm} }{\partial y} 
\pm\frac{1}{6}\frac{\partial^3V^{\pm} }{\partial y^3} = 0. 
\] 
 
Let consider the following initial conditions: 
\[ 
U(0,x)=0, \qquad Z(0,x)\sim\sum_{k\in \mathbb{Z} }Z_ke^{i\nu_kx}. 
\] 
We also assume that 
\[ 
 h(x)\sim\sum_{k\in \mathbb{Z} }h_ke^{i\mu_kx}, 
\] 
then the non-resonance condition can be written as  
\begin{equation} 
\label{math/23} 
\mu_k\ne \pm2\nu_l, \qquad \forall \; k,l \in \mathbb{Z}\quad \& \quad |k|+|l|\ne 0. 
\end{equation} 
In the periodic case, when the periods of functions $h(x)$ and $Z(0,x)$  
are equal to $2\pi$, we have that 
$\mu_k=\nu_k=k$. Thus, if $h(x)\ne \mbox{const}$ and $Z(0,x)\ne \mbox{const}$, 
then condition (\ref{math/23}) is not satisfied and system  
(\ref{math/20}) has a resonance.  
 
\subsection{Finite difference scheme} 
We define the space $\omega _h$ and time $\omega_{\tau}$ meshes and 
assume that the space mesh size $h$ and time mesh size $\tau$ are uniform. 
We denote by $v^n_j=v(t^n, y_j)$ a discrete function defined on  
$\omega_h \times \omega_\tau$. 
The following common notation of difference derivatives is used 
in our paper 
\begin{gather*} 
v_{\tau} = \frac{v^{n+1}-v^n} {\tau }, \qquad   
v_{\bar y} = \frac{v_j-v_{j-1}}{h},  \\  
 v_y = \frac{v_{j+1}-v_j}{h}, \qquad 
 v_{\buildrel \circ \over y} = \frac{v_{j+1}-v_{j-1}}{2h}. 
\end{gather*} 
The finite difference approximation of system (\ref{math/21})  
is defined as follows (see also \cite{furihata}): 
\begin{gather} 
  V_{\tau} = -\frac{1}{6}\left( \frac{V^{n+1} + V^n}{2} \right)_{\bar y y   
   \buildrel \circ \over y} 
 - \frac{3}{4} \left( \frac{\left(V^{n+1}\right)^2 + V^{n+1} V^n + \left(V^n\right)^2}{3} 
    \right)_{\buildrel \circ \over y} \nonumber\\ 
\qquad {} - \frac{F_{+}\left(W^{n+1}, W^n, j+1\right) - F_{+}\left(W^{n+1}, W^n, j-1\right)}{4h} , 
         \nonumber\\ 
   W_{\tau} = \frac{1}{6}\left( \frac{W^{n+1} + W^n}{2} \right)_{\bar y y  
   \buildrel \circ \over y} 
 + \frac{3}{4} \left( \frac{\left(W^{n+1}\right)^2 + W^{n+1} W^n + \left(W^n\right)^2}{3} 
    \right)_{\buildrel \circ \over y}\nonumber \\  
\qquad {} + \frac{F_{-}\left(V^{n+1}, V^n, j+1\right) - F_{-}\left(V^{n+1}, V^n, j-1\right)}{4h} ,\label{fds1} 
\end{gather} 
where the integrals are approximated as follows: 
\[ 
   F_{\pm}\left(V^{n+1}, V^n, j\right) = \frac{1}{2 \pi}\sum_{i=1}^N h(y_j \; {\mp}\; i\ h) 
    \frac{V^{n+1}_{j \mp 2i} + V^n_{j \mp 2i}}{2} \, h .    
\] 
The approximation  
error of this finite difference scheme is estimated as $O\left(\tau^2+h^2\right)$.   
Numerical methods for solving the Korteweg-de Vries equation are  
investigated in \cite{fornberg, sanz}.  
 
\subsection{Linear dispersion problem} \label{sec1}  
In this section we consider a linear problem 
\begin{gather} 
  Z_t+(H U)_x = - \frac{\varepsilon}{3} \;U_{xxx} , \nonumber\\  
U_t+Z_x = 0 .\label{math/29} 
\end{gather}  
First we will prove that system (\ref{math/29}) defines an ill-posed 
problem. Let consider the case $H = 1$. 
After simple computations we get the equation for $U$: 
\begin{equation} 
   U_{tt} - U_{xx} = \frac{\varepsilon}{3} \;U_{xxxx} . 
        \label{math/30} 
\end{equation} 
Considering the $k$-th Fourier mode we get that the solution of (\ref{math/30}) 
is unstable for $k^2 \varepsilon \geq 3$. In order to define a stable 
solution we use the following regularized problem   
\begin{gather} 
  Z_t+(H U)_x = - \frac{\varepsilon}{3} \;U_{xxx} 
   - \dfrac{\varepsilon^2}{20} \;U_{xxxxx},\nonumber \\  
  U_t+Z_x = 0.\label{math/31} 
\end{gather}  
 
We note that the averaged system (\ref{math/21}) also gives a nontrivial 
regularization of this ill-posed problem. 
 
\begin{figure}[th]
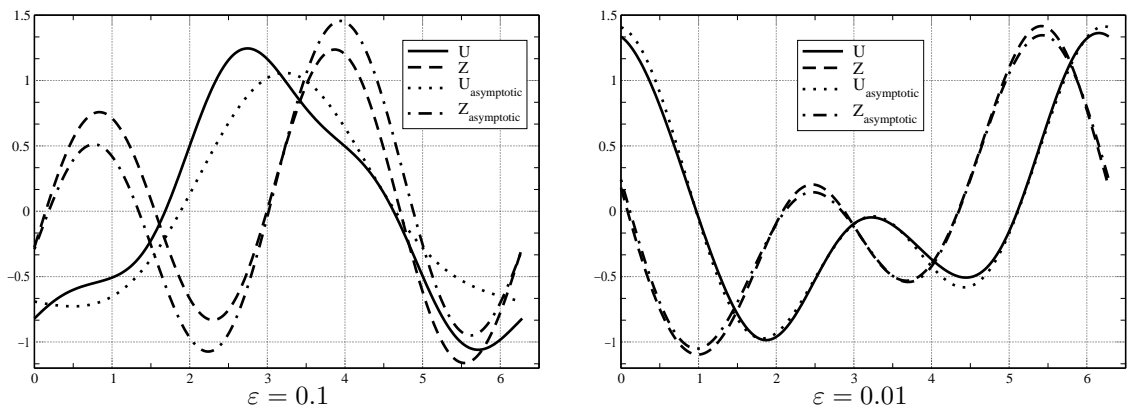
 
\begin{minipage}[b]{74mm} 
\hspace*{-2mm}\centerline{\rotatebox{270}{\epsfig{file=dis10.eps,width=50mm}}} 
 
\centerline{\small $\varepsilon=0.1$} 
\end{minipage} 
\hfill 
\begin{minipage}[b]{74mm} 
\centerline{\rotatebox{270}{\epsfig{file=dis100.eps,width=50mm}}} 
 
\centerline{\small $\varepsilon=0.01$} 
\end{minipage} 
 
\caption{Asymptotical solution for linear dispersion problem (\ref{math/29}).} 
\label{fig:d3} 
\end{figure}

The accuracy of asymptotic solution is illustrated by solving   
problem (\ref{math/31}) with following initial conditions 
\begin{equation}  
   U(x,0) = 0, \qquad Z(x,0) = \cos{x} + \sin{2 x}, \qquad 
   h(x) = 5 \sin{2 x}. 
   \label{math/32} 
\end{equation} 
Fig.~\ref{fig:d3} shows the solution of system (\ref{math/31}) and  
the asymptotic solution at $t= 1/\varepsilon$, for $\varepsilon = 0.1$ 
and $\varepsilon = 0.01$. 
For $\varepsilon = 0.001$ 
the difference between the asymptotic solution and the exact solution  
is too small to be illustrated on the figure. We also note, that  
the averaged system must be solved numerically only once and then 
a solution can be computed using simple interpolation procedure for  
any parameters $\varepsilon$.  
 
\vspace{6mm} 
 
\subsection{Nonlinear nondispersive problem}  
 
\begin{figure}[th]
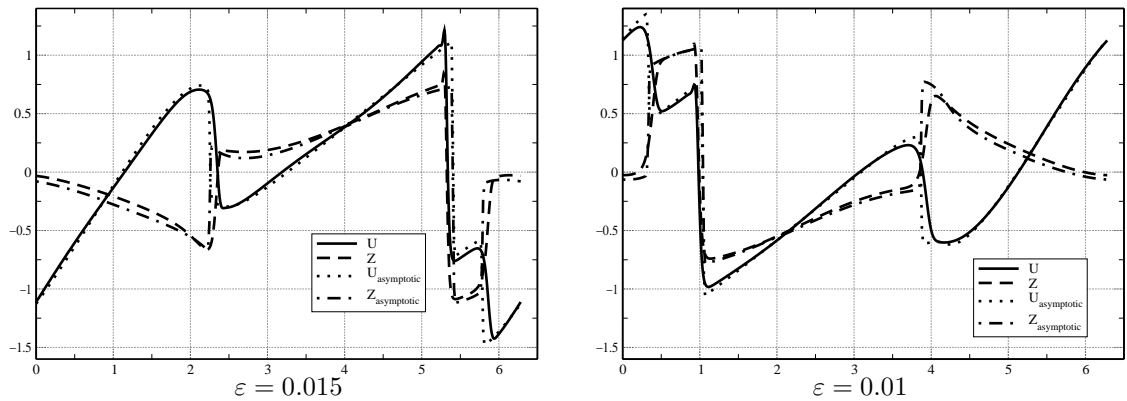
 
\begin{minipage}[b]{74mm} 
\hspace*{-2mm}\centerline{\rotatebox{270}{\epsfig{file=non66.eps,width=49mm}}} 
 
\centerline{\small $\varepsilon=0.015$} 
\end{minipage} 
\hfill \  
\begin{minipage}[b]{74mm} 
\centerline{\rotatebox{270}{\epsfig{file=non100.eps,width=49mm}}} 
 
\centerline{\small $\varepsilon=0.01$} 
\end{minipage} 
 
\caption{Asymptotical solution for nonlinear problem (\ref{math/33}).}\label{fig:n3} 
 
\end{figure}

In this section we consider a nonlinear problem 
\begin{gather} 
  Z_t+(HU)_x = - \varepsilon (ZU)_x , \nonumber\\ 
  U_t+Z_x = -\varepsilon U U_x  \label{math/33} 
\end{gather}  
with the same initial conditions (\ref{math/32}). As it follows from  
results given in previous sections, we have the resonance case.   
Fig.~\ref{fig:n3} shows the solution of system (\ref{math/33}) and 
the asymptotic solution at $t= 1/\varepsilon$.  
 
\vspace{6mm} 
 
\subsection{Shallow water waves} 
\begin{figure}[th]
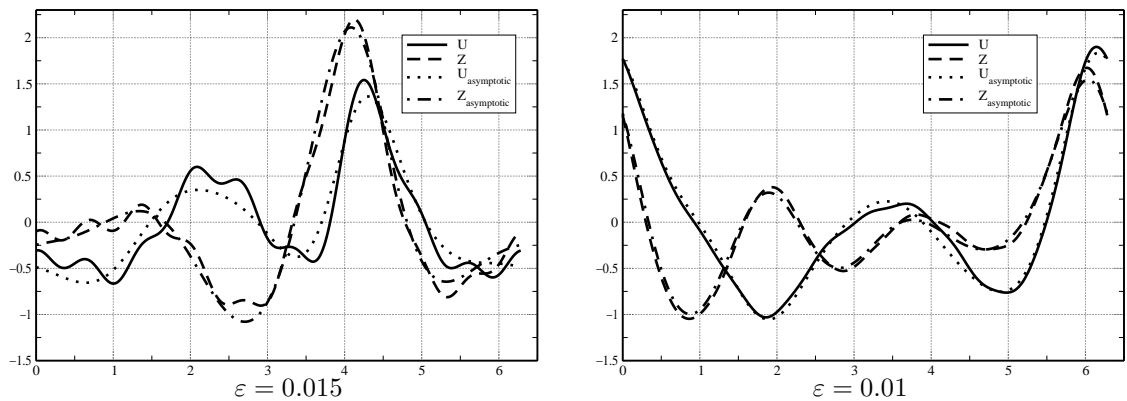
 
\begin{minipage}[b]{74mm} 
\hspace*{-2mm}\centerline{\rotatebox{270}{\epsfig{file=pilnas66.eps,width=49mm}}} 
 
\centerline{\small $\varepsilon=0.015$} 
\end{minipage} 
\hfill \  
\begin{minipage}[b]{74mm} 
\centerline{\rotatebox{270}{\epsfig{file=pilnas100.eps,width=49mm}}} 
 
\centerline{\small $\varepsilon=0.01$} 
\end{minipage} 
 
\caption{Asymptotical solution for the shallow water problem.}\label{fig:p3} 
 
\end{figure}

Here we solved the full nonlinear system  (\ref{math/18}). 
The regularization of section~\ref{sec1} is used in order to define 
the stable solution. 
Fig.~\ref{fig:p3} shows the solution of system (\ref{math/18}) and    
the asymptotic solution at $t= 1/\varepsilon$.  
 
\vspace{6mm} 
 
\subsection*{Acknowledgments} 
 
The authors are grateful to the anonymous referee for several 
useful comments and helpful remarks. 
 
\label{krylovas-lastpage} 
 

\begin{thebibliography}{99} 
\small 
 
\bibitem{Alesh} 
Aleshkov Ju~Z, Teorija Voln na Poverchnosti Tiazheloj Zhidkosti,  
Nauka~- Moscow, 1981 (in Russian). 
 
\bibitem{Bhat} 
Bhatnagar P~L, Nonlinear Waves in One-Dimensional Dispersive Systems, 
Oxford, 1979. 
 
\bibitem{boer} 
Boertjens G J and Van Horssen T, An Asymptotic Theory for a Weakly 
Nonlinear Beam Equation with a Quadratic Perturbation, {\it SIAM J. 
Appl. Math.} {\bf 60}, Nr.~2 (2000), 602--632.    
 
\bibitem{Bog} 
Bogoliubov N~N and Mitropolskii Ju A, Asymptotic Methods in the Theory of 
Non-Linear Oscillations, 
Gordon and Breach~- New York, 1961. 
 
\bibitem{ciegis} 
Krylovas A and {\v C}iegis R, Asymptotical Analysis of one   
Dimensional Gas Dynamics Equations, {\it Mathematical Modelling and 
Analysis} {\bf 6}, Nr.~1 (2001), 103--112. 
 
\bibitem{chikwendu} 
Chikwendu S C and Easwaran C V, A Laplace-Transform Multiple-Scale 
Procedure for the Asymptotic Solution of Weakly Nonlinear 
Partial Differential Equations, 
{\it Internat. J. Non-Linear Mech.} {\bf 34}, Nr.~1 (1999), 117--122. 
 
\bibitem{easwaran} 
Easwaran C V, A Scaled Characteristics Method for the 
Asymptotic Solution of Weakly Nonlinear Wave Equations, 
{\it Electronic Journal of Differential Equations} Nr.~3 (1998), 1--10.  
 
\bibitem{florence} 
Florence H and Denis S, Fast-Slow Dynamics for Parabolic 
Perturbations of Conservation Lows, 
{\it Commun. Part. Differ. Equat.} {\bf 21}, Nr.~9--10 (1996), 1587--1608.  
 
\bibitem{fornberg} 
Fornberg B and Driscoll T, A Fast Spectral Algorithm for Nonlinear 
Wave Equations with Linear Dispersion, 
{Journal of Comp. Physics} {\bf 155} (1999), 456--467. 
 
\bibitem{furihata} 
Furihata D, Finite Difference Schemes that Inherit Energy Conservation 
or Dissipation Property, {\it RIMS preprint Kyoto University}, Nr.~1212,  
1998. 
 
\bibitem{Jef} 
Jeffrey A and Kawahara T,  Asymptotic Method in Nonlinear Wave Theory, 
Boston, 1982. 
 
\bibitem{Kal1} 
Kaliakin L A, Long Wavelength Asymptotic. Integrability Equations as 
Asymptotic Limit of Nonlinear Systems,  
{\it Uspechi Matematicheskich Nauk} 
{\bf 44}, Nr.~1 (1989), 5--34 (in Russian). 
 
\bibitem{Cole} 
Kevorkian J and Cole J D, Multiple Scale and Singular  
Perturbation Methods, Springer-Verlag~- New-York, 1996. 
 
\bibitem{Kryl1} 
Krylov A V, About the Asymptotic Integration of First Order Hyperbolic Systems, 
{\it Lithuanian Mathematical Journal} {\bf 23}, Nr.~4 (1983), 12--17. 
 
\bibitem{Kryl2} 
Krylov A V, The Method of Research of Weakly Nonlinear Interaction One 
Dimensional Waves, {\it Prikladnaja Matematika i Mechanika} {\bf 51}, Nr.~6 (1987),  
933--940 (in Russian). 
 
\bibitem{Mitr} 
Mitropolskii Ju~A and Choma G P, About the Principle of Averaging Along the 
Characteristics for Hyperbolic Equations, 
{\it Ukrainskii Matematicheskii Zurnal} 
{\bf 22}, Nr.~5 (1970), 600--610 (in Russian). 
 
\bibitem{sanz} 
Sanz-Serna J M, Symplectic Integrators for Hamiltonian Problems, 
{\it Acta Numerica} {\bf 1} (1992), 243--286. 
 
\bibitem{lai} 
Shaoyoung L, The Asymptotic Theory for Semilinear Perturbed 
Telegraph Equation and its Application,  
{\it Appl. Math. and Mech. (Engl. ed.)} {\bf 18}, Nr.~7 (1997), 657--662.   
 
\bibitem{skinner} 
Skinner L A, Passages through Resonance in Weakly Nonlinear  
Systems, {\it IMA J. of Appl. Math.} {\bf 62} (1999), 45--60. 
 
\bibitem{Star} 
\v{S}taras A L, The Asymptotic Integration of Weakly Nonlinear Partial 
Derivatives Equations, {\em Doklady Akademii Nauk SSSR} {\bf 237}, Nr.~3 
(1977), 525--528 (in Russian). 
 
\bibitem{Tan} 
Taniuti T, Reductive Perturbation Method and Far Fields of Wave Equations, 
{\it Suppl. Progr. Theor. Phys.} {\bf 55} (1974), 1--35. 
\end{thebibliography}
\end{document}